\documentstyle[prb,aps,twocolumn,epsfig]{revtex}
\begin{document}
\draft
\twocolumn[\hsize\textwidth\columnwidth\hsize\csname @twocolumnfalse\endcsname\title{Temperature-dependent Raman spectroscopy in BaRuO$_3$ systems}
\title{Non-Fermi liquid behavior and scaling of low frequency suppression in
optical conductivity spectra of CaRuO$_3$}
\author{Y. S. Lee and Jaejun Yu}
\address{School of Physics and Center for Strongly Correlated Materials Research,\\
Seoul National University, Seoul 151-747, Korea}
\author{J. S. Lee and T. W. Noh}
\address{School of Physics and Research Center for Oxide Electronics, Seoul\\
National University, Seoul 151-747, Korea}
\author{T.-H. Gimm and Han-Yong Choi}
\address{School of Physics and Institute for Basic Science Research, Sung Kyun\\
Kwan University, Suwon 440-746, Korea}
\author{C. B. Eom}
\address{Department of Material Science and Engineering, University of\\
Wisconsin-Madison, Madison, Wisconsin 53706.}
\date{\today }
\maketitle

\begin{abstract}
Optical conductivity spectra $\sigma _1(\omega )$ of paramagnetic CaRuO$_3$
are investigated at various temperatures. At $T=10$ K, it shows a non-Fermi
liquid behavior of $\sigma _1(\omega )\sim 1/{\omega }^{\frac 12}$, similar
to the case of a ferromagnet SrRuO$_3$. As the temperature ($T$) is
increased, on the other hand, $\sigma _1(\omega )$ in the low frequency
region is progressively suppressed, deviating from the $1/{\omega }^{\frac 12%
}$-dependence. Interestingly, the suppression of $\sigma _1(\omega )$ is
found to scale with $\omega /T$ at all temperatures. The origin of the $%
\omega /T$ scaling behavior coupled with the non-Fermi liquid behavior is
discussed.
\end{abstract}


\vskip1pc] \newpage

The Fermi liquid model has provided a fundamental concept in understanding
metals.\cite{FL87} However, in some strongly correlated systems, non-Fermi
liquid (NFL) behaviors have been often observed, where the Fermi liquid
picture fails. In the normal state of high $T_c$ superconductors (HTS),
evidences of NFL behaviors were reported in many experiments involving
photoemission, transport, and optical measurements.\cite{orenstein00} In
particular, their optical conductivity spectra $\sigma _1(\omega )$ show $%
1/\omega $-dependence in contrast to the usual Drude form of $\sim 1/\omega
^2$, and their scattering rates show linear temperature- ($T$-) and $\omega $%
-dependences up to the mid-infrared (IR) region. Such unusual behaviors have
been explained in terms of the marginal Fermi liquid.\cite{Varma89}

Recently, perovskite ruthenates have attracted much attention as another
class of materials exhibiting NFL behavior. The ruthenates belong to 4$d$
transition metal oxides, and the electron correlation effects are believed
to play a crucial role in determining their physical properties.\cite{lee01}
Together with their intriguing transport properties,\cite{Allen96,Klein99}
the optical spectra show a distinct NFL behavior.\cite{schlesinger98,dodge00}
For an itinerant ferromagnet SrRuO$_3$ (a ferromagnetic transition
temperature $T_c$ = 165 K), Kostic {\it et al}.\cite{schlesinger98} reported
that its $\sigma _1(\omega )$ at low $T$ follow a $1/\omega ^{\frac 12}$%
-dependence, indicating a NFL behavior stronger than that of HTS. Recently,
Dodge {\it et al}.\cite{dodge00} fitted the $\sigma _1(\omega )$ of SrRuO$_3$
with $\sigma _\alpha (\omega )\sim (\tau ^{-1}-i\omega )^{-\alpha }$ with $%
\alpha \sim 0.4$ down to a very low energy region of $\sim 0.001$ eV.
Although the NFL behavior in the ruthenate has been widely accepted, its
origin is not clearly understood. In addition, an unusual suppression of $%
\sigma _1(\omega )$ in the low energy region occurs in the paramagnetic (PM)
state, but is absent in the ferromagnetic (FM) state.\cite{dodge00} However,
this intriguing phenomenon, which might be coupled with a NFL behavior, has
not been addressed properly.

While CaRuO$_3$ has the electronic structure similar to that of SrRuO$_3$,
this material does not show any magnetic ordering down to a very low $T$.%
\cite{Cao97} CaRuO$_3$ can provide a relatively wide $T$ window for
investigating the interesting PM state as well as another example for
understanding the NFL behavior. In this paper, we investigated the
electrodynamic responses of CaRuO$_3$. It was found that $\sigma _1(\omega )$
at 10 K follows $\sim 1/\omega ^{\frac 12}$, indicating a NFL behavior. With
increasing $T$, the suppression of $\sigma _1(\omega )$ near $\omega \simeq
0 $ develops below the characteristic energy $\omega _c$, which corresponds
to a peak structure in $\sigma _1(\omega )$ and shifts to higher frequencies
as $T$ increases. It is remarkable that the low frequency optical spectra in
a function of $\sigma _1(\omega )/\omega ^{-\frac 12}$ show a $\omega /T$
scaling behavior in a very wide $T$ range. While there have been similar
scaling behavior reported in some physical properties of other NFL systems,%
\cite{aronson95,Schroder98} the $\omega /T$ scaling behavior in ruthenates
is the first observation in optical spectra. This scaling indicates that the
only characteristic energy scale should be set by $T$ in the PM state of the
perovskite ruthenates.

\begin{figure}[tbp]
\epsfig{file=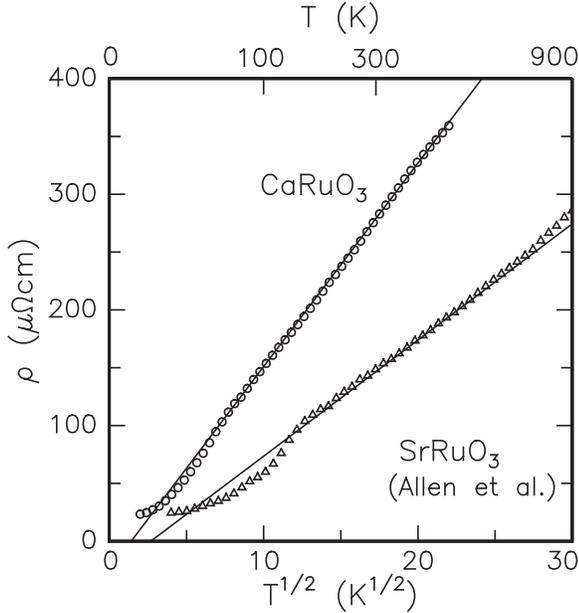,width=3.0in,clip=}
\vspace{2mm}
\caption{$T$-dependent $\rho (T)$ curves of a CaRuO$_3$ film (open circle)
and a SrRuO$_3$ single crystal (open triangle). The data of SrRuO$_3$ are
quoted from Ref. [5].}
\label{reflectivity}
\end{figure}

Several CaRuO$_3$ epitaxial films on (100) SrTiO$_3$ substrates were
fabricated using 90$^{\text{o}}$ off-axis sputtering techniques.\cite{Eom1}
Their thicknesses were about 5000 \AA . To obtain high crystalline quality
films with little strain effect, we used vicinal substrates with large
miscut angles (4 and 7$^{\text{o}}$). The $dc$ resistivity $\rho (T)$ was
measured up to 500 K using the standard four probe method. Figure 1 shows
the $\rho (T)$ curve of a film, which is nearly the same as that of a bulk
single crystal, including a crossover near around 50 K below which the $T$%
-dependence changes from $T^{\frac 12}$ to $T^{\frac 32}$.\cite{Cao97} The
300 K resistivity value of the film is $\sim $ 270 $\mu \Omega $cm,
comparable to that of the bulk single crystal value of resistivity $\sim $
200 $\mu \Omega $cm.\cite{Cao97} The resistivity ratio $\rho (300$ K)/$\rho
(10$ K) is about 9, indicating the high quality of our film. It is
interesting that the $\rho (T)$ in the PM state of the perovskite ruthenates
follows a $T^{\frac 12}$-dependence. The $\rho (T)$ of CaRuO$_3$ increases
continuously up to 500 K with no saturation, and $\rho (T)$ $\sim T^{\frac 12%
}$ above 50 K.\cite{Klein99} Note that, as displayed in Fig. 1, the reported 
$\rho $ values of a single crystal SrRuO$_3$ in the PM state also show the $%
T^{\frac 12}$-dependence.\cite{Allen96} The $T^{\frac 12}$-dependence of $%
\rho (T)$ in the perovskite ruthenates is another anomalous feature,
distinguished from the linear $T$-dependence of $\rho (T)$ in the normal
state of HTS.

Near normal incident reflectivity spectra $R(\omega )$ were measured in a
wide photon energy region of 5 meV $\sim $ 30 eV. The Kramers-Kronig (K-K)
analysis was used to calculate $\sigma _1(\omega )$ from the measured $%
R(\omega )$. For K-K transformation, $R(\omega )$ in the low frequency
region were extrapolated with the Hagen-Rubens relation. $T$-dependent $%
R(\omega )$ were measured in a photon energy region below 6 eV. Above 6 eV,
the room temperature $R(\omega )$ was used for high frequency extrapolation.
The overall features of the measured $R(\omega )$ were similar to those in
SrRuO$_3$ reported by Kostic {\it et al}..\cite{schlesinger98} The
calculated $\sigma _1(\omega )$ from the K-K analysis agreed with the
experimental $\sigma _1(\omega )$ independently obtained by spectroscopic
ellipsometry in the visible region, which demonstrates the validity of our
K-K analysis.\cite{yslee01} A high frequency region of $\sigma _1(\omega )$
in CaRuO$_3$ were described in our published paper.\cite{lee01} In the
paper, we focus on the far-IR region.

Figure 2 shows $T$-dependent $\sigma _1(\omega )$ in the far-IR region. The
peak at $\sim $ 570 cm$^{-1}$ is due to a transverse optic phonon mode,
whose $T$-dependence is rather weak. Interestingly, $\sigma _1(\omega )$ at
10 K shows a clear NFL behavior, deviating from that of conventional metals.
As shown in the inset of Fig. 2, $\sigma _1(\omega )$ at 10 K is
proportional to 1/$\omega ^{\frac 12}$,\cite{footnote1} which is much slower
than the frequency dependence of a Fermi liquid of 1/$\omega ^2$. Even at a
higher $T$, the 1/$\omega ^{\frac 12}$-dependence in $\sigma _1(\omega )$ is
retained in the high frequency region, which might be correlated with the $%
T^{\frac 12}$-dependence of $\rho (T)$ at high temperatures. A similar NFL
behavior was also observed in SrRuO$_3$.\cite{schlesinger98} It is
interesting that the 1/$\omega ^{\frac 12}$-dependence in $\sigma _1(\omega
) $ can be observed in perovskite ruthenates with different magnetic ground
states.

\begin{figure}[tbp]
\epsfig{file=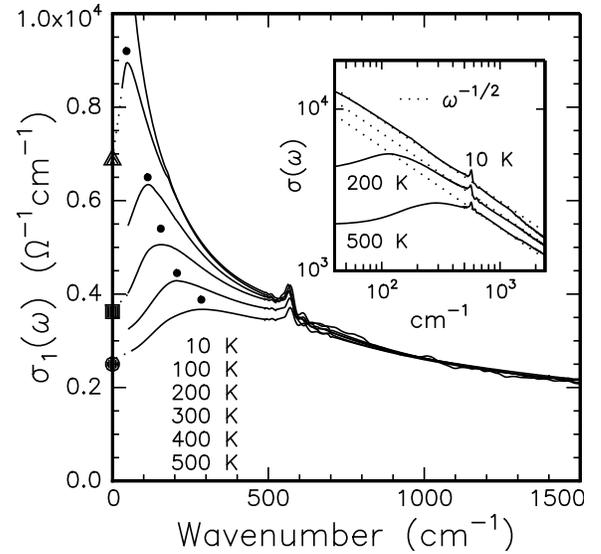,width=3.0in,clip=}
\vspace{2mm}
\caption{$T$-dependenct $\sigma _1(\omega )$ of CaRuO$_3$ below 1500 cm$%
^{-1} $. The solid circle symbols represent $\omega _c$. The solid triangle,
the solid square, and the solid circle symbols on the $y$-axis represent the 
$dc$ value of $\sigma (T)$ at 100, 300, and 500 K, respectively. The dotted
lines are $\sigma _1(\omega )$ obtained from the Hagen-Rubens
extrapolations. In the inset, $\sigma _1(\omega )$ at 10, 200, 500 K follow
1/$\omega ^{\frac 12}$ above $\omega _c$. For clarity, the $\sigma _1(\omega
)$ at 200 K and 500 K are multiplied by a factor of 0.85 and 0.7,
respectively. The dotted lines are guidelines for 1/$\omega ^{\frac 12}$%
-dependence.}
\label{conductivity1}
\end{figure}

The suppression in $\sigma _1(\omega )$ is observed near $\omega \simeq 0$
at higher $T$. As $\omega $ decreases from the high frequency side, $\sigma
_1(\omega )$ increase initially but decrease below the peak frequency $%
\omega _c$, approaching smoothly to the measured {\it dc} conductivity
values. This feature is clear even at 100 K, where the {\it dc} conductivity
value is rather high by $\sim $ 7000 $\Omega ^{-1}$cm$^{-1}$. It is noted
that in the case of SrRuO$_3$, the low frequency suppression occurs only in
its PM state, not in its FM state.\cite{dodge00} Together with the $T^{\frac 
12}$-dependence of $\rho (T)$, the low-energy suppression in $\sigma
_1(\omega )$ can be regarded as a generic feature of the PM state of the
perovskite ruthenates.

Note that the low-energy suppression of $\sigma _1(\omega )$ shows an
interesting $T$-dependent evolution. As $T$ increases, the suppression
feature becomes clear and $\omega _c$ shifts to a higher frequency linearly
with $T$. It is evident that the peak structure does not arise from
electronic transitions or disorder effects. The values of $\omega _c$ are
comparable with a thermal energy, $k_BT$. This energy scale is too low for a
typical interband transition.\cite{lee01} A similar suppression in $\sigma
_1(\omega )$ near $\omega \simeq 0$ has been often observed in highly
disordered systems, but their characteristic energy scale is expected to
decrease with increasing $T$,\cite{basov98} which is the opposite to our
case. Therefore, a thermal energy scale of a pseudogap-like feature observed
in CaRuO$_3$ is quite unique.

The pseudogap-like feature could be closely related to nearly ferromagnetic
characteristics of CaRuO$_3$. Several experimental and theoretical evidences
suggest that CaRuO$_3$ should be nearly ferromagnetic.\cite
{Yoshimura99,cava01,Mazin97} A strong FM fluctuation was also observed in
the PM state of SrRuO$_3$.\cite{Yoshimura99} Especially, local density
functional calculations on (Ca,Sr)RuO$_3$ showed that the lattice
distortions associated with different ionic sizes are crucial in determining
the magnetic properties.\cite{Mazin97} Further, phonon anomalies at $T_c$
were observed in SrRuO$_3$, indicating the strong spin-lattice interaction.%
\cite{Iliev99} These imply that the lattice degree of freedom is strongly
coupled to the magnetic ordering so that the excitation of a relevant phonon
mode can be responsible for the local magnetic fluctuation in the nearly FM
system. From the spin fluctuation theory of nearly ferromagnetic materials,%
\cite{moriya} it is known that the mean-square amplitude of spin-fluctuation
increases linearly in proportion to $T$. Thus, one may expect the low-energy
quasiparticle excitations to be strongly renormalized by the thermally
induced spin-fluctuations with spin-lattice coupling, where such $T$%
-dependent renormalization might be relevant to the suppression of $\sigma
_1(\omega )$ in the low frequency region. It is noted that the
pseudogap-like feature in $\sigma _1(\omega )$ of CaRuO$_3$ and its
proximity to the FM instability is quite analogous to the situation in HTS,
which is close to the antiferromagnetic instability.\cite{Manske01}

Now, we show that from the systematic $T$-dependent evolution of the low
frequency suppression in $\sigma _1(\omega )$, an interesting $\omega /T$
scaling behavior can occur in the perovskite ruthenates. The low frequency
suppression in $\sigma _1(\omega )$ can be expressed as a deviation from the
1/$\omega ^{\frac 12}$-dependence. As shown in the inset of Fig. 2, the
deviation region of $\sigma _1(\omega )$ from 1/$\omega ^{\frac 12}$%
-dependence becomes wider with increasing $T$, consistent with the shift of $%
\omega _c$ to higher frequency. To check the possibility of scaling
behavior, we plotted $\sigma _1(\omega )/a\omega ^{-\frac 12}$ vs. $\omega
/T $. With the value of the coefficient $a$ adopted for the scaling, the $%
\sigma _1(\omega )$ at 10 K was reproduced. As shown in Fig. 3, all of the
normalized conductivity spectra collapse onto a single line.\cite{footnote3}
It is noted that this scaling behavior persists up to a rather high
temperature, 500 K. The $\omega /T$ scaling behavior means that the $T$%
-dependent suppression behavior could be determined only by $T$, indicating

\begin{equation}
\sigma _1(\omega )\sim 1/\omega ^{\frac 12}\cdot Z(\omega /T)\text{,}
\end{equation}
or 
\begin{equation}
\sigma _1(\omega )T^{\frac 12}\sim (T/\omega )^{\frac 12}\cdot Z(\omega /T)%
\text{,}
\end{equation}
The scaling function $Z(\omega /T)$ is fitted quite well with $Z(\omega
/T)=\tanh (\beta \omega /T)$, with $\beta =1.6$. Clearly, Eq. (1) and (2)
are closely related to the characteristic properties in the PM states, such
as $\rho (T)\sim T^{\frac 12}$ and $\sigma _1(\omega )\sim 1/\omega ^{\frac 1%
2}$ at high frequencies. We also plotted the SrRuO$_3$ $\sigma _1(\omega )$
data in the PM region (i.e., at 185 K, 225 K, and 250 K) reported by Kostic 
{\it et al}..\cite{schlesinger98} Interestingly, the normalized spectra of
SrRuO$_3$ fall on the scaling curve. This indicates that the scaling
function shown in Fig. 3 could be applied to other perovskite ruthenates.

\begin{figure}[tbp]
\epsfig{file=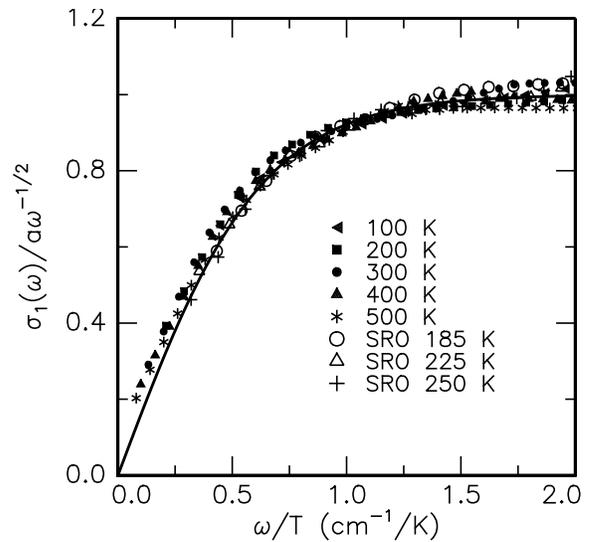,width=3.0in,clip=}
\vspace{2mm}
\caption{$T$-dependent $\sigma _1(\omega )/a\omega ^{-\frac 12}$ with $%
\omega /T$ as the abscissa. The open circle, open trigonal, and cross
symbols are for the 185 K, 225 K, and 250 K spectra of SrRuO$_3$,
respectively, quoted from Ref. [7]. For SrRuO$_3$, the value of $a$ was
adopted to reproduce the 40 K $\sigma _1(\omega )$. The solid line
represents $Z(\omega /T) = \tanh(1.6\omega/T)$. }
\label{scaling behavior}
\end{figure}

While the $\omega /T$ scaling behavior in $\sigma _1(\omega )$ of the
perovskite ruthenates is quite unique, it is noted that similar
scale-invariance has also been observed in some physical properties of other
NFL systems, such as some $f$-electron compounds\cite{aronson95} and HTS.%
\cite{Schroder98} The scaling behaviors in these NFL systems indicate that
the only characteristic energy scale is determined by $T$, a possible origin
of which was suggested to be the quantum critical fluctuation associated
with the zero temperature phase transition.\cite{Sachdev99} Similar to other
NFL\ systems, our $\omega /T$ scaling behavior in $\sigma _1(\omega )$ may
suggest a quantum critical point between the ferromagnetic and paramagnetic
phases in the perovskite ruthnates: The ferromagnetic transition temperature 
$T_c$ is decreased as $x$ is increased in Sr$_{1-x}$Ca$_x$RuO$_3$ and is
completely suppressed in CaRuO$_3$, and a quantum critical point is expected
at an appropriate value of $x=x_c$.\cite{Cao97,Fukunaga94} We note that,
consistent with the magnetic quantum phase transition, the previous low $T$
transport measurements hinted a phase transition from a Fermi liquid
behavior for SrRuO$_3$ to a NFL behavior for CaRuO$_3$.\cite
{Fukunaga94,Capogna02} Motivated by our observation, understanding the
origin of the NFL behavior associated with the $\omega /T$ scaling in $%
\sigma _1(\omega )$ and its possible relation with the quantum criticality
is a challenging issue in the future study of perovskite ruthenates.

In summary, the optical spectra of the nearly ferromagnetic CaRuO$_3$ shows
non-Fermi liquid behavior and a scaling in the low frequency suppression.
Its $\sigma _1(\omega )$ follow the $1/\omega ^{\frac 12}$-dependence,
similar to the case of SrRuO$_3$. From the $T$-dependent evolution of the
low frequency suppression, it is observed that the $\sigma _1(\omega )$
normalized by the $1/\omega ^{\frac 12}$ can be scaled with $\omega /T$ at
all temperatures, indicating that only characteristic energy scale is
determined by $T$. The $\omega /T$ scaling coupled with the non-Fermi liquid
behavior is expected to provide further insights into understanding the
unusual electrodynamics of the ruthenates.

We thank Yunkyu Bang, J. E. Han, In-Sang Yang, Philip B. Allen, J. H. Kim,
and K. H. Kim for helpful discussions. We also thank PAL and NCIRF for
allowing us to use some of their facilities. This work was supported by
Ministry of Science and Technology through the Creative Research Initiative
program, and by KOSEF through CSCMR. THG and HYC acknowledge the support
from KOSEF through Grant No.\ 1999-2-11400-005-5.


\begin{references}
\bibitem{footnote1}  From the extended Drude model analysis, it is found
that the frequency-dependent scattering rate at 10 K follows the $\omega ^{%
\frac 12}$-dependence with the correlation to the $1/\omega ^{\frac 12}$%
-dependence in $\sigma _1(\omega )$.
\bibitem{footnote3}  It is not clear that even the 10 K $\sigma _1(\omega )$
follows the scaling behavior. If so, this might be observed below 20 cm$^{-1}
$, where is beyond the experimental region.
\end{references}
\end{document}